\documentclass[11pt,twoside]{article}
\usepackage[latin1]{inputenc}
\usepackage{amssymb}
\usepackage{amsmath}
\usepackage{exscale}
\usepackage[dvips]{graphicx}

\catcode`@=11
\renewcommand{\section}{\@startsection{section}{1}{\z@}{\medskipamount}
{\medskipamount}{\large\bf}}
\numberwithin{equation}{section}
\catcode`@=12

\begin{document}

\begin{titlepage}
\setcounter{page}{0}
\begin{flushright}
ITP--UH--05/02
\end{flushright}
\vspace*{2.0cm}
\begin{center}
{\Large\bf Interaction of Noncommutative Plane Waves\\[2mm]
in 2+1 Dimensions}\\
\vspace*{14mm}
{\large Stig Bieling}\\[0.5cm]
{\em Institut für Theoretische Physik \\
Universität Hannover \\
Appelstraße 2, 30167 Hannover, Germany}\\
{Email: bieling@itp.uni-hannover.de}
\end{center}
\vspace{2cm}
\begin{abstract}
\noindent
In this paper the interaction of extended waves in a
noncommutative modified 2+1 dimensional $U(2)$ sigma model are
studied. Using the dressing method, we construct an explicit two-wave
solution of the noncommutative field equation. 
The scattering of these waves and large time factorization are
discussed.
\end{abstract}
\vfill
\end{titlepage}

%---------------------------------------------------------
%---------------introducion-------------------------------
%---------------------------------------------------------
\section{Introduction}

Possible noncommutativity of spacetime coordinates is an old idea \cite{Snyder:1946qz}
which was recently revived in active studying of noncommutative field theories (see e.g. reviews
\cite{Nekrasov:2000ih}\cite{Douglas:2001ba}\cite{Konechny:2001wz}\cite{Szabo:2001kg}
and references therein). Being nonlocal, these theories have many interesting
properties.
\medskip

It has been found that noncommutative field theories
arise naturally in certain limits of string theories involving $D$-branes 
and a nontrivial $B$-field background \cite{Seiberg:1999vs}.
Recently, this fact led to a high level of interest in field theories
on noncommutative spaces and their nonperturbative solutions (like solitons), 
which admit a $D$-brane interpretation
in the context of string theory 
\cite{Gopakumar:2000zd}\cite{Nekrasov:2000ih}\cite{Harvey:2001yn}\cite{Douglas:2001ba}. 
A lot of progress has been made in the analysis of such solitonic solutions of 
various noncommutative field theories (see e.g. 
\cite{Nekrasov:2000ih}\,-\,\cite{Konechny:2001wz},
\cite{Gopakumar:2000zd}\,-\,\cite{Furuta:2002ty}  and references therein).
\medskip

In \cite{Ooguri:1991fp} it was shown that the open $N=2$ 
fermionic string is identical at tree level
to self-dual Yang-Mills theory in 2+2 dimensions.
Turning on a constant $B$-field in this string theory yields 
\textit{noncommutative} self-dual Yang-Mills (ncSDYM) \cite{Lechtenfeld:2000nm}. 
If the $B$-field is restricted to the
world volume of $n$ coincident $D2$-branes the effective field theory turns out
to be the noncommutative generalization \cite{Lechtenfeld:2001uq} of an integrable 
modified $U(n)$ sigma model in 2+1 dimensions introduced by Ward \cite{Ward:ie}. 
Its soliton solutions have been studied in \cite{Lechtenfeld:2001aw}.
\medskip

In the present paper we examine a different type of solutions of the noncommutative
modified sigma model mentioned above, namely plane waves, which were studied in the 
commutative setup by Leese \cite{Leese:hj}. Our work is based on \cite{Lechtenfeld:2001aw}, 
where the gauge group is restricted to the case of $U(2)$. The organization is the following. 
After presenting the necessary material and formulas for the commutative sigma model 
we briefly review its noncommutative extension and the general construction of solutions. 
We then derive an explicit two-wave solution and investigate its scattering properties
and large time factorization.

%---------------------------------------------------------
%---------------U(2) modell-------------------------------
%---------------------------------------------------------
\section{Noncommutative modified $U(2)$ sigma model}

The investigations are based on a modified $SU(2)$ sigma model in 2+1 dimensions
introduced by Ward \cite{Ward:ie}. This model describes the dynamics of an $SU(2)$-valued
field $J(t,x,y)$ depending on the real coordinates $t,x,y$ of $\mathbb{R}^{1,2}$ with
minkowskian signature $(-\,+\,+)$. Its field equation reads
\begin{equation}\label{sigma1}
\eta^{\mu\nu}\partial_{\mu}\left(J^{-1}\partial_{\nu}J\,\right)+
v_{\alpha}\epsilon^{\alpha\mu\nu}\partial_{\mu}\left(J^{-1}\partial_{\nu}J\,\right)=0,
\end{equation}
with a constant unit vector $(v_{\alpha})=(0,1,0)$. It is obvious
that the second term in (\ref{sigma1}) breaks Lorentz invariance, but makes
this equation integrable \cite{Ward:ie}.
The model has a positive definite energy functional, which
is given by 
\begin{equation}\label{energy}
E=\int dxdy\,{\cal E}=\frac{1}{2}\int dxdy\,\mbox{tr}\left(\partial_tJ^{-1}\partial_tJ+
\partial_xJ^{-1}\partial_xJ+\partial_yJ^{-1}\partial_yJ\right).
\end{equation}
\vspace{0.5mm}

The field equation (\ref{sigma1}) actually arises from the Bogomol'nyi equations
for the 2+1 dimensional Yang-Mills-Higgs system \cite{Ward}
\begin{equation}\label{ymh1}
\frac{1}{2}\,\epsilon_{\alpha\mu\nu}{\cal F}^{\mu\nu}=\partial_\alpha\Phi+
[{\cal A}_{\alpha},\Phi\,].
\end{equation}
After choosing the gauge ${\cal A}_x=-\Phi$ and ${\cal A}_t={\cal A}_y$ and solving two of the 
three equations above by
${\cal A}_x=\frac{1}{2}\,J^{-1}\partial_xJ$ and ${\cal A}_t=\frac{1}{2}\,J^{-1}
\left(\partial_tJ+\partial_yJ\right)$ one obtains
\begin{equation}\label{ymh2}
\partial_x\left(J^{-1}\partial_xJ\right)+\partial_y\left(J^{-1}\partial_yJ\right)-
\partial_t\left(J^{-1}\partial_tJ\right)+\partial_y\left(J^{-1}\partial_tJ\right)-
\partial_t\left(J^{-1}\partial_yJ\right)=0,
\end{equation}
which is precisely equation (\ref{sigma1}). It admits both soliton \cite{Ward:ie}
and extended wave solutions \cite{Leese:hj}.
\medskip

The noncommutative generalization of equation (\ref{sigma1}) is easily achieved by
deforming the ordinary product of functions into the noncommutative star product, 
which is given by
\begin{equation}\label{star1}
(f\star g)(x)=f(x)\exp[\frac{i}{2}\,\overleftarrow{\partial}_{\mu}
\,\theta^{\mu\nu}\,\overrightarrow{\partial}_{\nu}]g(x).
\end{equation}
Since the time coordinate remains commutative, the only nonvanishing components of
the tensor $\theta^{\mu\nu}$ are
\begin{equation}\label{thetacom}
\theta^{xy}=-\theta^{yx}=:\theta>0.
\end{equation}
After introducing the following combinations of the coordinates $t$ and $y$
\begin{equation}\label{uv}
u:=\frac{1}{2}\,(t+y)\:\:,\:\:v=\frac{1}{2}\,(t-y)\:\:\Rightarrow\:\:
\partial_u=\partial_t+\partial_y\:\:,\:\:\partial_v=\partial_t-\partial_y,
\end{equation}
the noncommutative field equation\footnote{It should be noted here, that in the
noncommutative setup $J\in U(2)$, since $SU(2)$ as gauge group is not possible.
See for example \cite{Matsubara:2000gr}\cite{Armoni:2000xr}.} reads \cite{Lechtenfeld:2001uq}
\begin{equation}\label{sigma2}
\partial_x\left(J^{-1}\star\partial_xJ\right)-\partial_v\left(J^{-1}\star\partial_uJ\right)=0,
\end{equation}
and the energy density is 
\begin{eqnarray}\label{nc-en}
{\cal E}&=&\frac{1}{2}\,\mbox{tr}\,\left[\partial_tJ^{\dag}\star\partial_tJ
+\partial_xJ^{\dag}\star\partial_xJ+\partial_yJ^{\dag}
\star\partial_yJ\right].
\end{eqnarray}
\vspace{0.05cm}

The nonlocality of the star product involves difficulties in explicit
calculations. It is therefore more convenient to switch to the operator formalism.
Here, the star product is traded for operator-valued spatial coordinates $\hat{x},\hat{y}$,
which are subject to the following commutation relations
\begin{equation}\label{com}
[t,\hat{x}]=[t,\hat{y}]=0\:\:,\:\:[\hat{x},\hat{y}]=i\theta.
\end{equation}
With the complex coordinates, $\hat{z}=\hat{x}+i\hat{y}\:,\:
\hat{\bar{z}}=\hat{x}-i\hat{y},$ relation (\ref{com}) becomes
$[\hat{z},\hat{\bar{z}}]=2\theta$. Now one can introduce creation and annihilation operators
\begin{equation}\label{aa+}
a^{\dag}=\frac{1}{\sqrt{2\theta}}\,\hat{\bar{z}}\:\:,\:\:
a=\frac{1}{\sqrt{2\theta}}\,\hat{z}\:\:\Rightarrow\:\:[a,a^{\dag}]=1,
\end{equation}
which act in a harmonic oscillator Hilbert space ${\cal H}$ with orthonormal basis
$\left\{|n\rangle,n=0,1,2,\ldots\right\}$ in standard fashion
\begin{equation}\label{ortho}
a|n\rangle=\sqrt{n}\,|n-1\rangle\:\:,\:\:a^{\dag}|n\rangle=\sqrt{n+1}\,|n+1\rangle.
\end{equation}
Every c-number function $f(t,z,\bar{z})$ can be associated with an operator-valued function
$\hat{f}\equiv F(t,a,a^{\dag})$ through the Moyal-Weyl map
\begin{eqnarray}\label{weyl}
f(t,z,\bar{z})\rightarrow F(t,a,a^{\dag})
&=&\frac{1}{(2\pi)^2}\,\int
d^2k\tilde{f}(k_x,k_y)e^{-i(k_x\hat{x}+k_y\hat{y})}\nonumber\\
&=&-\int
\frac{dpd\bar{p}}{(2\pi^2)}\,dzd\bar{z}f(t,z,\bar{z})e^{-i[\bar{p}\,(\sqrt{2\theta}\,a-z)+p\,
(\sqrt{2\theta}\,a^{\dag}-\bar{z})]},
\end{eqnarray}
with $p=\frac{1}{2}\,(k_x+ik_y)$. The inverse transformation reads
\begin{equation}
f(t,z,\bar{z})=F_{\star}\left(t,\frac{z}{\sqrt{2\theta}},\frac{\bar{z}}{\sqrt{2\theta}}\right),
\end{equation}
where $F_{\star}$ is obtained by replacing ordinary products with star products.
Under the Moyal-Weyl map one has \cite{Harvey:2001yn}\cite{Morita:2000fs}
\begin{equation}\label{weyl2}
\partial_z\rightarrow\hat{\partial}_z=-\frac{1}{\sqrt{2\theta}}\,[a^{\dag},\:\:]\:,
\hspace{2cm}\partial_{\bar{z}}\rightarrow\hat{\partial}_{\bar{z}}=\frac{1}{\sqrt{2\theta}}
\,[a,\:\:]\:,
\end{equation}
\begin{equation}
f\star g\rightarrow\hat{f}\hat{g}\:,\hspace{2cm}
\int dxdyf=2\pi\theta\mbox{Tr}_{{\cal H}}\hat{f}=2\pi\theta\sum_{n\geq0}
\langle n|\hat{f}|\,n\rangle.
\end{equation}
Hence, in terms of creation and annihilation operators the energy density (\ref{nc-en})
becomes
\begin{equation}\label{nc-en1}
\hat{{\cal E}}=\frac{1}{2}\,\mbox{tr}\,\left[\partial_t\hat{J}^{\dag}
\partial_t\hat{J}\right]+
\frac{1}{2\theta}\,\mbox{tr}\,\left[
[a,\hat{J}^{\dag}][a,\hat{J}^{\dag}]^{\dag}+
[a^{\dag},\hat{J}^{\dag}][a^{\dag},\hat{J}^{\dag}]^{\dag}\right], \end{equation}
where $\hat{\partial}_x=\hat{\partial}_z+\hat{\partial}_{\bar{z}}$ and 
$\hat{\partial}_y=i\left(\hat{\partial}_z-\hat{\partial}_{\bar{z}}\right)$ have been used,
and $\hat{J}\equiv J(t,a,a^{\dag})$. 
For simplicity, hats over operators are omitted from now on.

%---------------------------------------------------------
%-----------------dressing approach-----------------------
%---------------------------------------------------------
\section{Construction of solutions}

The main observation for constructing solutions of the equation of motion (\ref{sigma2}) is
that this equation can be formulated as the compatibility condition of a system of linear 
differential equations. For the commutative model the `dressing method' 
\cite{zak}\cite{Forgacs:1983gr} can be used to construct solitons and plane waves as shown in 
\cite{Ward:ie} and \cite{Leese:hj}. This can easily be extended to the
noncommutative situation, which will be briefly presented next. 
For more details see \cite{Lechtenfeld:2001aw}.
\medskip

Let us consider the following linear system
\begin{eqnarray}\label{lin}
\left(\zeta\partial_x-\partial_u\right)\psi&=&A\psi,\nonumber\\
\left(\zeta\partial_v-\partial_x\right)\psi&=&B\psi,
\end{eqnarray}
where $\psi=\psi(x,u,v,\zeta)$ is a $2\times2$ matrix whose elements are operators acting in
${\cal H}$. The matrices $A$ and $B$ are of the same type but do not depend on the parameter
$\zeta\in\mathbb{C}$. The matrix $\psi$ has to satisfy the reality condition \cite{Ward:ie}
\begin{equation}\label{real}
\psi(x,u,v,\zeta)\left[\psi(x,u,v,\bar{\zeta}\right]^{\dag}=1.
\end{equation}
The integrability conditions for the system (\ref{lin}) are
\begin{equation}\label{intcon}
\partial_xB-\partial_vA=0\:\:\:,\:\:\:\partial_xA-\partial_uB-\left[A,B\right]=0.
\end{equation}
The second equation can be solved by parametrizing the matrices $A$ and $B$ in the following 
way :
\begin{equation}\label{intcon2}
A=J^{-1}\partial_uJ\:\:\:,\:\:\:B=J^{-1}\partial_xJ,
\end{equation}
where $J$ is some unitary $2\times2$ matrix. Then the first equation in (\ref{intcon}) becomes
\begin{equation}\label{intcon3}
\partial_x\left(J^{-1}\partial_xJ\right)-\partial_v\left(J^{-1}\partial_uJ\right)=0,
\end{equation}
which is the operator version of the field equation (\ref{sigma2}).
Thus, any solution of the system (\ref{lin}) yields a solution of (\ref{intcon3}) since
\begin{equation}\label{j}
\psi(x,u,v,\zeta=0)=J^{-1}(x,u,v).
\end{equation}
\vspace{0.05cm}

Let us now assume that $\psi$ is of the form 
\cite{zak}\cite{Forgacs:1983gr}\cite{Ward:ie}\cite{Lechtenfeld:2001aw}
\begin{equation}\label{ansatz}
\psi=1+\sum_{p=1}^m\frac{R_p}{\zeta-\mu_p}\hspace{1cm},\hspace{1cm}
R_p=\sum_{l=1}^mT_l\Gamma^{lp}T^{\dag}_p,
\end{equation}
where $\mu_p$ are complex constants, the $T_k(t,x,y)$ are some $2\times1$ matrices
and the $\Gamma^{kl}$ are some operator-valued functions. By multiplying (\ref{lin})
with $\psi^{-1}$ on the right and using (\ref{real}), the linear system can be rewritten as
\begin{eqnarray}\label{lin2}
-\psi(\zeta)\left(\zeta\partial_x-\partial_u\right)
\psi(\bar{\zeta})^{\dag}&=&A,\nonumber\\
-\psi(\zeta)\left(\zeta\partial_v-\partial_x\right)
\psi(\bar{\zeta})^{\dag}&=&B.
\end{eqnarray}
The singularities at $\zeta=\mu$ and $\zeta=\bar{\mu}$ have to be removable, since
the right hand side of these equations does not depend on $\zeta$. By putting to zero the
corresponding residues one finds two differential equations 
\begin{eqnarray}\label{residuum1}
&\left(1-\sum_{p=1}^m\frac{R_p}{\mu_p-\bar{\mu}_k}\right)\left(\bar{\mu}_k\partial_x
-\partial_u\right)R_k^{\dag}=0,&\nonumber\\
&\left(1-\sum_{p=1}^m\frac{R_p}{\mu_p-\bar{\mu}_k}\right)\left(\bar{\mu}_k\partial_v
-\partial_x\right)R_k^{\dag}=0.&
\end{eqnarray}
In the same manner, one obtains from (\ref{real})
\begin{equation}\label{residuum2}
\left(1-\sum_{p=1}^m\frac{R_p}{\mu_p-\bar{\mu}_k}\right)T_k=0, \end{equation}
which yields the inverse of the functions $\Gamma^{kl}$ in terms of
the matrices $T_k$ \cite{Lechtenfeld:2001aw},
\begin{equation}\label{inverse}
\tilde{\Gamma}_{kl}=\frac{T_k^{\dag}T_l}{\mu_k-\bar{\mu}_l}\hspace{1cm}\mbox{and}
\hspace{1cm}\sum^m_{p=1}\Gamma^{lp}\,\tilde{\Gamma}_{pk}=\delta^l_k.
\end{equation}
\vspace{0.05cm}

It is useful to introduce the following new coordinates
\begin{eqnarray}\label{movecoord}
w_k&:=&\nu_k(x+\bar{\mu}_ku+\frac{1}{\bar{\mu}_k}\,v)=
\nu_k[x+\frac{1}{2}\,(\bar{\mu}_k-\frac{1}{\bar{\mu}_k})y+
\frac{1}{2}\,(\bar{\mu}_k+\frac{1}{\bar{\mu}_k})t],\nonumber\\
\bar{w}_k&:=&\bar{\nu}_k(x+\mu_ku+\frac{1}{\mu_k}\,v)=
\bar{\nu}_k[x+\frac{1}{2}\,(\mu_k-\frac{1}{\mu_k})y+
\frac{1}{2}\,(\mu_k+\frac{1}{\mu_k})t],
\end{eqnarray}
with the $\nu_k$ being functions of the $\mu_k$,
\begin{equation}\label{nu}
\nu_k=\left[\frac{4i}{\mu_k-\bar{\mu}_k-\mu_k^{-1}+\bar{\mu}^{-1}_k}\,
\frac{\mu_k-\mu_k^{-1}-2i}{\bar{\mu}_k-\bar{\mu}^{-1}_k+2i}\,\right]^{1/2}.
\end{equation}
Since
\begin{equation}\label{kom2}
[w_k,\bar{w}_k]=2\theta,
\end{equation}
one can define the `co-moving' creation and annihilation operators
\begin{equation}\label{co}
c_k=\frac{1}{\sqrt{2\theta}}\,w_k\:\:,\:\:c_k^{\dag}=\frac{1}{\sqrt{2\theta}}\,\bar{w}_k\:\:
\Rightarrow\:[c_k,c_k^{\dag}]=1,
\end{equation}
with oscillator basis
$\left\{\,|n\rangle_k\right\}$. Derivatives with respect to these coordinates are
represented in the same way as in (\ref{weyl2}),
\begin{equation}\label{cc+}
\sqrt{2\theta}\,\partial_{w_k}=-[c_k^{\dag},\:\:
]\:\:,\:\:\sqrt{2\theta}\,\partial_{\bar{w}_k}=[c_k,\:\:]\:.
\end{equation}
\vspace{0.05cm}

One can relate the co-moving oscillators to the static one.
Expressing the $w_k$ through $z$ and $\bar{z}$, one obtains 
\begin{equation}\label{c-a}
c_k=(\cosh\tau_k)\,a-(e^{i\vartheta_k}\sinh\tau_k)\,a^{\dag}-\beta_kt,
\end{equation}
where $\hspace{0.2cm}\nu_k=\cosh\tau_k-e^{i\vartheta_k}\sinh\tau_k$\hspace{0.2cm},\hspace{0.2cm}
$e^{i\vartheta_k}\tanh\tau_k
=\frac{\bar{\mu}_k-\bar{\mu}_k^{-1}-2i}{\bar{\mu}_k-\bar{\mu}_k^{-1}+2i}\hspace{0.2cm}$ and
\begin{equation}\label{constants}
\beta_k:=-\frac{1}{2}\,(2\theta)^{-1/2}\,\nu_k(\bar{\mu}_k+\bar{\mu}_k^{-1}).
\end{equation}
Equations (\ref{residuum1}) can now be combined to a single one,
\begin{equation}\label{residuum3}
\left(1-\sum_{p=1}^m\frac{R_p}{\mu_p-\bar{\mu}_k}\right)\,c_kT_k=0.
\end{equation}
With (\ref{residuum2}) it is quite obvious that a sufficient condition for
a solution is
\begin{equation}\label{Z_k}
c_k\,T_k=T_k\,Z_k
\end{equation}
with some operator $Z_k$. If $Z_k$ is taken to be $c_k$ then (\ref{Z_k}) becomes
\begin{equation}\label{Z_k2}
[c_k,T_k]=0,
\end{equation}
which means that any matrix $T_k$ with elements being arbitrary functions of $c_k$
yields a solution $\psi$ of the linear system (\ref{lin}), and by (\ref{j}) a solution
$J$ of the equation of motion (\ref{intcon3}).

%---------------------------------------------------------
%------------------two wave system -----------------------
%---------------------------------------------------------
\section{A two-wave configuration}

As was shown in the previous section, the solution for $J^{-1}=J^{\dag}$ can
be expressed through $R_k$ 
\begin{equation}\label{general}
J^{\dag}=1-\sum^m_{k=1}\frac{R_k}{\mu_k}\,.
\end{equation}
In order to obtain an extended wave solution comparable to the commutative solution of Leese,
 the elements of the $T_k$ are taken to be
exponentials of $b_kw_k=b_k\,\sqrt{2\theta}c_k$ \footnote{Compare with \cite{Leese:hj}.},
 with $b_k\in\mathbb{R}$. To be specific
\begin{equation}\label{T_k}
T_k=\left(\begin{array}{c}
            1\\
            e^{b_k\omega_k}\end{array}\right).
\end{equation}
From now on let us take $m=2$ in formula (\ref{general}),
\begin{equation}\label{m=2}
J^{\dag}=1-\frac{1}{\mu_1}\,T_1\Gamma^{11}T_1^{\dag}-
\frac{1}{\mu_2}\,T_1\Gamma^{12}T_2^{\dag}
-\frac{1}{\mu_1}\,T_2\Gamma^{21}T_1^{\dag}-\frac{1}{\mu_2}\,T_2\Gamma^{22}T_2^{\dag}.
\end{equation}
Now the functions $\Gamma^{kl}$ can be expressed in terms of $\tilde{\Gamma}_{lp}$ via
(\ref{inverse})
\begin{eqnarray}
\Gamma^{11}&=&[\tilde{\Gamma}_{11}-\tilde{\Gamma}_{12}\tilde{\Gamma}_{22}^{-1}
\tilde{\Gamma}_{21}]^{-1}\ ,\nonumber\\
\Gamma^{12}&=&-[\tilde{\Gamma}_{11}-\tilde{\Gamma}_{12}\tilde{\Gamma}_{22}^{-1}
\tilde{\Gamma}_{21}]^{-1}
\tilde{\Gamma}_{12}\tilde{\Gamma}_{22}^{-1}\ ,\nonumber\\
\Gamma^{21}&=&-[\tilde{\Gamma}_{22}-\tilde{\Gamma}_{21}\tilde{\Gamma}_{11}^{-1}
\tilde{\Gamma}_{12}]^{-1}
\tilde{\Gamma}_{21}\tilde{\Gamma}_{11}^{-1}\ ,\nonumber\\
\Gamma^{22}&=&[\tilde{\Gamma}_{22}-\tilde{\Gamma}_{21}\tilde{\Gamma}_{11}^{-1}
\tilde{\Gamma}_{12}]^{-1}\ ,
\end{eqnarray}
which yields $J^{\dag}$ in explicit $T_k$-dependence
\begin{eqnarray}\label{final}
J^{\dag}&=&1-T_1[\,T_1^{\dag}(1-\sigma
P_2\,)\,T_1]^{-1}T_1^{\dag}(\,\frac{\mu_{11}}{\mu_1}-\frac{\mu_{21}}{\mu_2}
\,\sigma\,P_2\,)\nonumber\\
& &-T_2[\,T_2^{\dag}(1-\sigma
P_1\,)\,T_2]^{-1}T_2^{\dag}(\,\frac{\mu_{22}}{\mu_2}-\frac{\mu_{12}}{\mu_1}\,\sigma\,P_1\,),
\end{eqnarray}
where the following complex parameters have been introduced
\begin{equation}\label{const2}
\mu_{kl}=\mu_k-\bar{\mu}_l\:\:,\:\:\sigma=\frac{\mu_{11}\mu_{22}}{\mu_{12}\mu_{21}}\,.
\end{equation}
The $P_k$ are hermitian projectors parametrized through the matrices $T_k$,
\begin{equation}\label{projek}
P_k=T_k(T_k^{\dag}T_k)^{-1}T_k^{\dag}\:\:,\:\:P_k^{\dag}=P_k\:\:,\:\:P_k^2=P_k.
\end{equation}
For simplicity, we choose the first wave to be static,
$\mu_1=-i$, which means $w_1=z$, and rename $\mu_2:=\mu\:,\:w_2:=w.$ In addition, $\mu$ is
restricted to be pure imaginary, $\mu=ip$, with real $p>1$. This ensures
the existence of large time limits of $\exp{(b_2w)}$ and $\exp{(b_2\bar{w})}$, which will be
important in further considerations. The strategy is to evaluate the energy density ${\cal E}$
for large positive and large negative times and investigate the interaction.
\medskip

For $\mu=ip$ the constant $\beta$ in (\ref{c-a}) will be real
\begin{equation}\label{beta}
\beta=-\frac{1}{2\sqrt{\theta}}\,\frac{p-\frac{1}{p}}{\sqrt{p+\frac{1}{p}}}<0,
\end{equation}
so that large-time limits can be easily evaluated. For $t$ being large means\footnote{$b_2$
is restricted to be greater than zero, in order to achive a well defined limit.}
$b_2w=b_2\sqrt{2\theta}c\rightarrow+b_2\sqrt{2\theta}|\beta| t$, and finally
$e^{b_2w}\rightarrow+\infty\:,e^{b_2\bar{w}}\rightarrow+\infty$.
In this limit, $J^{\dag}$ becomes\footnote{We refrain from writing down the
full expression for $J^{\dag}$.} $(b:=b_1)$
\begin{eqnarray}\label{t++}
J^{\dag}_{+\infty}&:=&\lim_{t\rightarrow+\infty}J^{\dag}\nonumber\\
&=&1-2\left(\begin{array}{cc}
[1+(1-\sigma)e^{b\bar{z}}e^{bz}]^{-1} &
-\frac{1+p}{1-p}\,[1+(1-\sigma)e^{b\bar{z}}e^{bz}]^{-1}e^{b\bar{z}} \\
\frac{1+p}{1-p}\,[1+(1-\sigma)e^{bz}e^{b\bar{z}}]^{-1}e^{bz} &
[1+(1-\sigma)e^{bz}e^{b\bar{z}}]^{-1}\end{array}\right).\\
\nonumber
\end{eqnarray}
For large negative times,
$e^{b_2w}\rightarrow0\:,e^{b_2\bar{w}}\rightarrow0,$ and we obtain $(b:=b_1)$
\begin{eqnarray}\label{t--}
J^{\dag}_{-\infty}&:=&\lim_{t\rightarrow-\infty}J^{\dag}\nonumber\\
&=&-1-2\left(\begin{array}{cc}
-(1-\sigma)[1-\sigma+e^{b\bar{z}}e^{bz}]^{-1} &
\frac{1+p}{1-p}\,[1-\sigma+e^{b\bar{z}}e^{bz}]^{-1}e^{b\bar{z}} \\
-\frac{1+p}{1-p}\,[1-\sigma+e^{bz}e^{b\bar{z}}]^{-1}e^{bz} &
-(1-\sigma)[1-\sigma+e^{bz}e^{b\bar{z}}]^{-1}\end{array}\right),\\
\nonumber
\end{eqnarray}
where the first sign results from simplifying matrix elements. Now the energy density
can be readily calculated.
\medskip
  
For large times, $J\rightarrow J_0+{\cal O}(t^{-1}),$ with $J_0$ being independent of $t$,
and therefore in this limit the first term in equation (\ref{nc-en1}) vanishes.
Hence, the asymptotics of the energy density
operator are\footnote{To avoid any confusion, operators are marked again with hats.}
\begin{equation}\label{nc-en11}
\hat{{\cal E}}_{\pm\infty}:=\lim_{t\rightarrow\pm\infty}\hat{{\cal E}}=
\frac{1}{2\theta}\,\mbox{tr}\,\left[
[a,\hat{J}^{\dag}_{\pm\infty}][a,\hat{J}^{\dag}_{\pm\infty}]^{\dag}+
[a^{\dag},\hat{J}^{\dag}_{\pm\infty}][a^{\dag},\hat{J}^{\dag}_{\pm\infty}]^{\dag}\right],
\end{equation}
with $\hat{J}^{\dag}_{\pm\infty}$ given by (\ref{t++}) and (\ref{t--}).
By noting that $e^{b\hat{\bar{z}}}e^{b\hat{z}}\rightarrow e^{2bx-b^2\theta}$ and
that the star product of two functions depending only on $x$ is identical to
usual multiplication, one sees that the final result in star product formulation is

\begin{eqnarray}\label{result}
{\cal E}_{-\infty,\star}&=&4b^2\left\{\mbox{sech}^2
\left[b\,x-\frac{b^2\theta}{2}+\gamma\right]
+\mbox{sech}^2\left[b\,x+\frac{b^2\theta}{2}+\gamma\right]\right\},\nonumber\\
{\cal E}_{+\infty,\star}&=&4b^2\left\{\mbox{sech}^2
\left[b\,x-\frac{b^2\theta}{2}-\gamma\right]
+\mbox{sech}^2\left[b\,x+\frac{b^2\theta}{2}-\gamma\right]\right\},
\end{eqnarray}
with $\gamma$ being a real constant depending only on $p$,
\begin{equation}\label{phase}
\gamma=-\frac{1}{2}\,\ln{\left(\frac{1+p}{1-p}\right)^2}.
\end{equation}

These two noncommutative plane waves interact in a simple way (see Fig. 1-4 
on the next page).
The phase of the static wave depends on the noncommutativity parameter 
$\theta$. This wave experiences a phase shift of $2\gamma$, which is 
independent of $\theta$ and precisely matches the result
obtained from the commutative model by Leese\footnote{
See \cite{Leese:hj} and formula (4.5) for $p_1=-1$ and $p_2=p$.}.
The phase shift is clearly visible in the figures on the next page.
Moreover, it is easy to see that in the commutative limit, $\theta\rightarrow0$,
the formulas (\ref{result}) recover the energy density as obtained by Leese\footnote{
See \cite{Leese:hj} and formula (4.4).}.

\begin{figure}[ht]
\begin{minipage}[h]{7.5cm}
\begin{center}
\includegraphics[height=4cm,width=7cm]{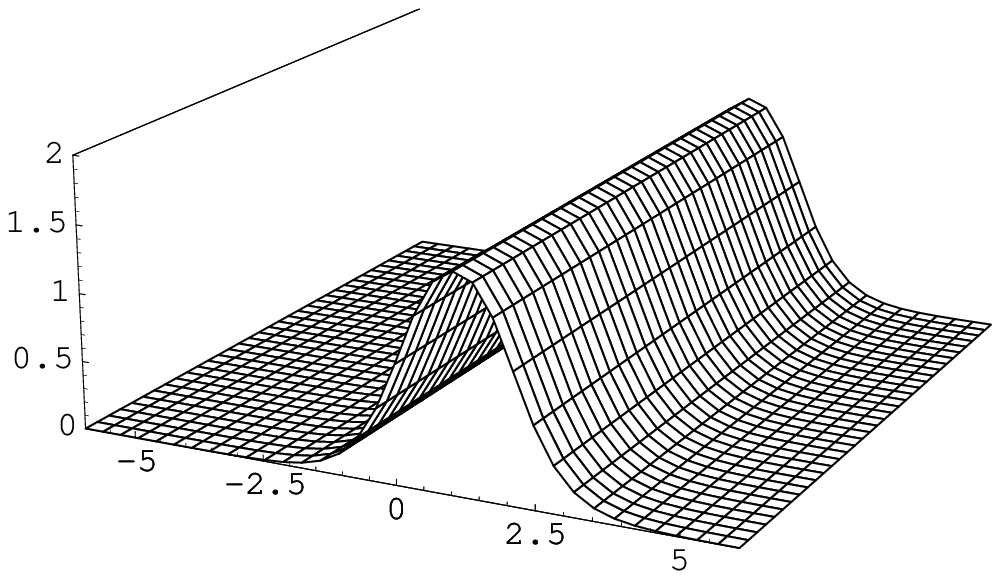}
\caption{${\cal E}_{-\infty,\star}$ with $b=1,p=2,\theta=1$.}
\end{center}
\end{minipage}
\hfill
\begin{minipage}[h]{7.5cm}
\begin{center}
\includegraphics[height=4cm,width=7cm]{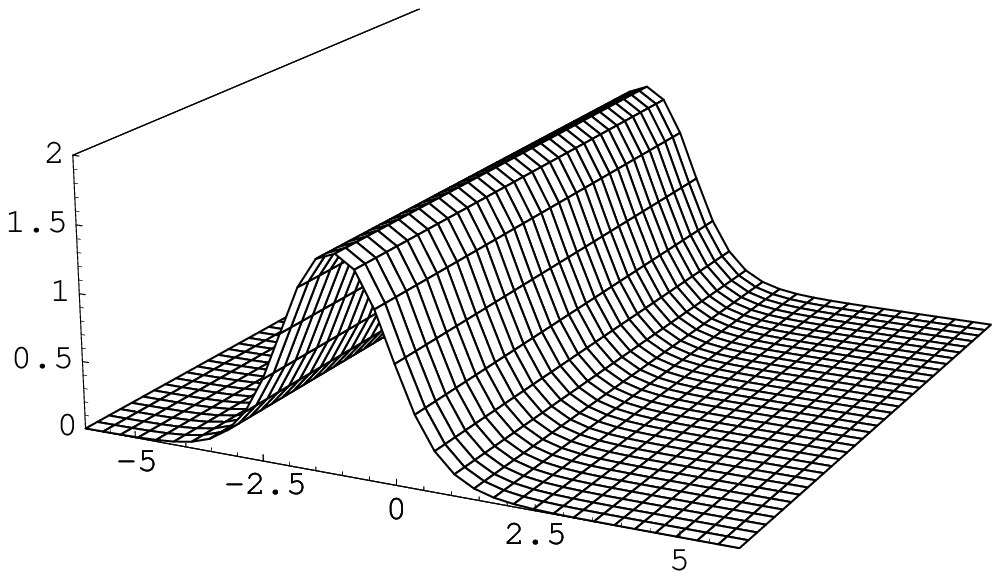}
\caption{${\cal E}_{+\infty,\star}$ with $b=1,p=2,\theta=1$.}
\end{center}
\end{minipage}
\end{figure}
\begin{figure}[ht]
\begin{minipage}[h]{7.5cm}
\begin{center}
\includegraphics[height=4cm,width=7cm]{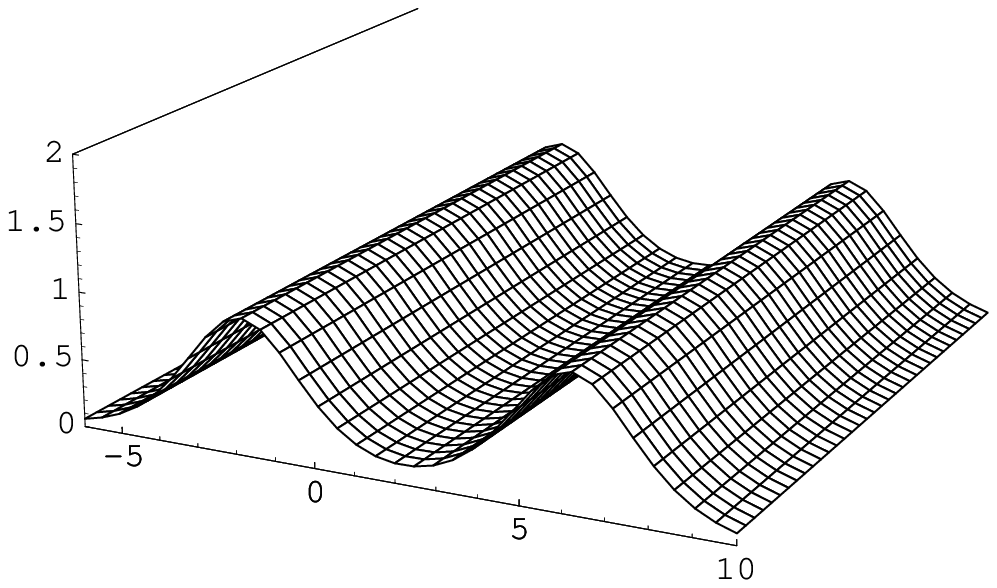}
\caption{${\cal E}_{-\infty,\star}$ with $b=1,p=2,\theta=4$.}
\end{center}
\end{minipage}
\hfill
\begin{minipage}[h]{7.5cm}
\begin{center}
\includegraphics[height=4cm,width=7cm]{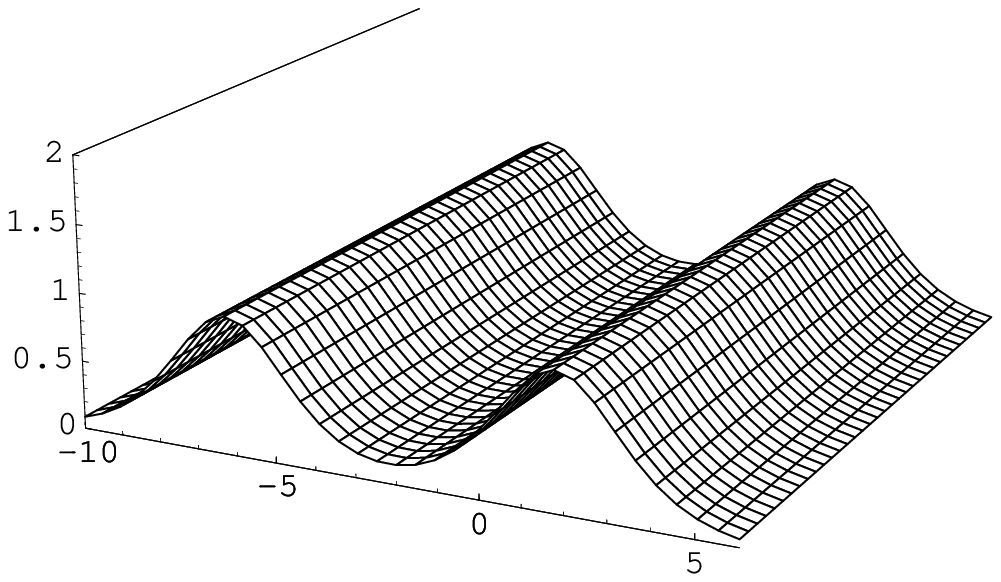}
\caption{${\cal E}_{+\infty,\star}$ with $b=1,p=2,\theta=4$.}
\end{center}
\end{minipage}
\end{figure}

%--------------------------------------------------------
%----------------faktorisierung--------------------------
%--------------------------------------------------------

\section{Large time factorization}

Besides the additive ansatz used in (\ref{m=2}), one can also use a multiplicative ansatz. 
Let us take the function $\hat{\psi}(t,\hat{x},\hat{y},\zeta)$ to be 
of the form
\begin{equation}
\hat{\psi}_2=\left(1+\frac{-2i}{\zeta+i}\,\tilde{P}_1\right)
\left(1+\frac{\mu-\bar{\mu}}{\zeta-\mu}\,\tilde{P}_2\right),
\end{equation}
where the operators $\tilde{P}_1$ and $\tilde{P}_2$ are yet to be determined. 
Because the reality condition
(\ref{real}) has to hold, $\tilde{P}_1$ and $\tilde{P}_2$ are identified as hermitian
projectors, parametrized as usual, $\tilde{P}_k=\tilde{T}_k
(\tilde{T}^{\dag}_k\tilde{T}_k)^{-1}\tilde{T}_k^{\dag}\:\:,\linebreak[4]k=1,2$, with some
$2\times 1$ matrices $\tilde{T}_k.$ The removability of the singularities at $\zeta=\bar{\mu}$
and $\zeta=\mu$
in (\ref{lin2}) is guaranteed if $\tilde{P}_2$ satisfies the equation
\begin{equation}\label{pole1}
(1-\tilde{P}_2)c\tilde{P}_2=0\hspace{1cm}\Rightarrow\hspace{1cm}
c\,\tilde{T}_2=\tilde{T}_2Z_2,
\end{equation}
which means, for $Z_2=c$, that the elements of $\tilde{T}_2$ are functions of $c$.
For the residues at $\zeta=\pm i$, one obtains the equation
\begin{equation}\label{pole2}
\frac{1}{\sqrt{2\theta}}\,(1-\tilde{P}_1)\left(1+\frac{\mu-\bar{\mu}}{i-\mu}\,
\tilde{P}_2\right)a\left(1+\frac{\bar{\mu}-\mu}{i-\bar{\mu}}\,\tilde{P}_2\right)\tilde{P}_1=0.
\end{equation}
This equation looks more complicated than (\ref{pole1}), but can be investigated in large 
positive and large negative time limit, yielding an equation similiar to (\ref{pole1}).
Let $\tilde{T}_2$, solving equation (\ref{pole1}), be of the form
\begin{eqnarray}\label{tilde1}
\tilde{T}_2=\left(\begin{array}{c}
                 \gamma_1\\\gamma_2e^{\gamma_3c}
                 \end{array}\right),
\end{eqnarray}
with constant parameters $\gamma_k\:,\:k=1,2,3\,,$ and $\gamma_3>0$.
Then, for $t\rightarrow\pm\infty,$ one gets
\begin{eqnarray}\label{inf1}
\lim_{t\rightarrow-\infty}\tilde{P}_2=
                                \left(\begin{array}{cc}
                                1 & 0 \\ 0 & 0
                                \end{array}\right)=:\Pi_{-\infty}\hspace{.7cm},\hspace{.7cm}
\lim_{t\rightarrow+\infty}\tilde{P}_2=
                                \left(\begin{array}{cc}
                                0 & 0 \\ 0 & 1
                                \end{array}\right)=:\Pi_{+\infty}.
\end{eqnarray}
So, in these limits $\tilde{P}_2$ is a constant projector. Therefore, the operator $a$ in
equation (\ref{pole2}) can be moved to the left of $\tilde{P}_1$ and (\ref{pole2}) simplifies 
to $(\tilde{P}_{1,\pm\infty}:=\lim_{t\rightarrow\pm\infty}\tilde{P}_1)$
\begin{equation}\label{inf2}
(1-\tilde{P}_{1,\pm\infty})a\tilde{P}_{1,\pm\infty}=0\hspace{1cm}\Rightarrow\hspace{1cm}
a\,\tilde{T}_{1,\pm\infty}=\tilde{T}_{1,\pm\infty}Z_1.
\end{equation}
With $Z_1=a$, the elements of $\tilde{T}_{1,\pm\infty}$ are arbitrary functions of $a$,
and in this case $\tilde{T}_{1,\pm\infty}$ can be chosen to be
\begin{eqnarray}\label{tilde2}
\tilde{T}_{1,-\infty}=\left(\begin{array}{c}
                 \lambda_1\\\lambda_2e^{\lambda_3a}
                 \end{array}\right)\hspace{1cm},\hspace{1cm}
\tilde{T}_{1,+\infty}=\left(\begin{array}{c}
                 \lambda_4\\\lambda_5e^{\lambda_6a}
                 \end{array}\right),
\end{eqnarray}
where the $\lambda_k$ are yet to be determined. Writing
\begin{equation}\label{fak1}
\hat{J}_{\pm\infty}^{\dag}=\lim_{t\rightarrow\pm\infty}\hat{\psi}_2(\zeta=0)=
(1-2\tilde{P}_{1,\pm\infty})(1-2\Pi_{\pm\infty}),
\end{equation}
and comparing (\ref{tilde2}) with formulas (\ref{t++}) and (\ref{t--}), one finds
\begin{eqnarray}\label{T+-}
\tilde{T}_{1,-\infty}=\left(\begin{array}{c}
                 \lambda\\e^{b\hat{z}}
                 \end{array}\right)=\left[1+(\lambda-1)
\Pi_{-\infty}\right]T_1\hspace{.5cm},\hspace{.5cm}
\tilde{T}_{1,+\infty}=\left(\begin{array}{c}
                 1\\ \lambda e^{b\hat{z}}
                 \end{array}\right)=\left[1+(\lambda-1)
\Pi_{+\infty}\right]T_1,
\end{eqnarray}
where $\lambda:=\frac{1+p}{1-p}$ and $T_1$ is given in (\ref{T_k}).
So, $\hat{J}^{\dag}$ factorizes in the large time limit in a term $(1-2\tilde{P}_{1,\pm\infty})$
constructed from a modified time dependent one-wave solution
$(T_1\rightarrow\tilde{T}_{1,\pm\infty})$ and a constant unitary
matrix $(1-2\Pi_{\pm\infty})=:U^{\dag}_{\pm\infty}.$ The energy density $\hat{{\cal
E}}_{\pm\infty}$ from equation (\ref{nc-en1}) now reads
\begin{eqnarray}\label{nc-en2}
\hat{{\cal E}}_{\pm\infty}&=&\frac{1}{2\theta}\,\mbox{tr}\,\Big[
[a,(1-2\tilde{P}_{1,\pm\infty})U^{\dag}_{\pm\infty}][a,(1-2\tilde{P}_{1,\pm\infty})
U^{\dag}_{\pm\infty}]^{\dag}\nonumber\\
& &+\mbox{}[a^{\dag},(1-2\tilde{P}_{1,\pm\infty})U^{\dag}_{\pm\infty}]
[a^{\dag},(1-2\tilde{P}_{1,\pm\infty})U^{\dag}_{\pm\infty}]^{\dag}\Big]\nonumber\\
&=&\frac{2}{\theta}\,\mbox{tr}\,\Big[
[a,\tilde{P}_{1,\pm\infty}][a,\tilde{P}_{1,\pm\infty}]^{\dag}
+\mbox{}[a^{\dag},\tilde{P}_{1,\pm\infty}][a^{\dag},\tilde{P}_{1,\pm\infty}]^{\dag}\Big],
\end{eqnarray}
where the projectors $\tilde{P}_{1,\pm\infty}$ are given by
\begin{eqnarray}\label{enproj}
\tilde{P}_{1,-\infty}&=&\left(\begin{array}{cc}
(1-\sigma)[(1-\sigma)+e^{b\hat{\bar{z}}}e^{b\hat{z}}]^{-1} &
\frac{1+p}{1-p}\,[(1-\sigma)+e^{b\hat{\bar{z}}}e^{b\hat{z}}]^{-1}e^{b\hat{\bar{z}}} \\
\frac{1+p}{1-p}\,[(1-\sigma)+e^{b\hat{z}}e^{b\hat{\bar{z}}}]^{-1}e^{b\hat{z}} &
1-[(1-\sigma)+e^{b\hat{z}}e^{b\hat{\bar{z}}}]^{-1}\end{array}\right),\nonumber\\
\nonumber\\
\nonumber\\
\tilde{P}_{1,+\infty}&=&\left(\begin{array}{cc}
[1+(1-\sigma)e^{b\hat{\bar{z}}}e^{b\hat{z}}]^{-1} &
\frac{1+p}{1-p}\,[1+(1-\sigma)e^{b\hat{\bar{z}}}e^{b\hat{z}}]^{-1}e^{b\hat{\bar{z}}} \\
\frac{1+p}{1-p}\,[1+(1-\sigma)e^{b\hat{z}}e^{b\hat{\bar{z}}}]^{-1}e^{b\hat{z}} &
(1-\sigma)\left\{1-[1+(1-\sigma)e^{b\hat{z}}e^{b\hat{\bar{z}}}]^{-1}\right\}\end{array}\right).
\end{eqnarray}
By construction the operators $\hat{{\cal E}}_{\pm\infty}$ from
(\ref{nc-en2}) are identical to those in equation (\ref{nc-en1})
with $\hat{J}^{\dag}_{\pm\infty}$ given by (\ref{t++}) and
(\ref{t--}). So, both additive and multiplicative ans\"atze result in
the same energy densities ${\cal
E}_{\pm\infty,\star}$ given by formulas (\ref{result}).

%---------------------------------------------------------
%---------------concluding remarks------------------------
%---------------------------------------------------------

\section{Conclusion}

In this paper we have constructed an explicit time-dependent two-wave solution of a 
noncommutative 2+1 dimensional modified $U(2)$ sigma model. We have shown that this 
configuration depends on the noncommutativity parameter $\theta$, but the phase shift
produced by the interaction is independent of $\theta$, coinciding with the commutative
case. We have also proven that this solution factorizes in the large time limit.
\medskip

As was previously remarked, the considered noncommutative modified sigma model 
can be obtained from ncSDYM in 2+2 dimensions \cite{Lechtenfeld:2000nm}.
For the commutative case it is well known that most integrable equations 
in three or less dimensions derive from SDYM by suitable dimensional reductions
\cite{Ward:gz}-\cite{Dimakis:2000tm}.  
It therefore will be interesting to consider other reductions of ncSDYM 
(see e.g. \cite{Paniak:2001fc}\,-\,\cite{Legare:2000pb}\,) and to
construct solutions of those noncommutative integrable equations with the 
help of the dressing approach.

\section*{Acknowledgements}

I would like to thank Alexander~D.~Popov for useful comments and discussions. This work was 
done within the framework of the DFG priority program (SPP 1096) in string theory.

%---------------------------------------------------------
%-----------------------literatur-------------------------
%---------------------------------------------------------

\end{document}